\documentclass[twocolumn, amsmath, superscriptaddress, amsfonts,prb]{revtex4-2}
\usepackage{graphicx}
\usepackage{url}
\pdfoutput=1
\usepackage{mhchem}

\begin{document}

\title{Electron-hole liquid in biological tissues under ultra high dose rate ionizing radiation}

\author{Diana Shvydka}\email{diana.shvydka@osumc.edu}\affiliation{Department of Radiation Oncology, The Ohio State University Wexner Medical Center, Columbus, OH 43221, USA}
\author{Victor Karpov}\email{victor.karpov@utoledo.edu}\affiliation{Department of Physics and Astronomy, University of Toledo, Toledo,OH 43606, USA}

\begin{abstract}
We develop a quantitative model of ionization processes in biological tissues under Ultra High Dose Rate (UHDR) radiation. The underlying conjecture is that of electron-hole liquid (EHL) forming in water based substances of biological tissues. Unlike the earlier known EHL in semiconductor crystals, the charge carriers here are low mobile due to strong interactions with the background (solvated electrons, etc.); hence, EHL resembling ionic melts. Similar to all ionic systems, the Coulomb coupling makes that EHL energetically favorable that leads to recombination barriers suppressing subsequent structural transformations. In particular, generation of secondary reactive species in such EHL becomes limited translating into reduction of  biological damages and tissue sparing effect. We show how these processes are sensitive to the tissue quality and frequency dispersion of the dielectric permittivity. Equations for dose and dose rate defining the sparing thresholds are derived.

\end{abstract}

\maketitle

\section{Introduction}\label{sec:intro}

A seminal paper of 2014 \cite{favaudon2014} triggered intensive quest into understanding the Ultra High Dose Rate (UHDR) radiation effects in biological tissues. Nowadays, these effects are considered a new modality in radiation treatment (RT) of cancer known as FLASH RT. Its unique feature is a reduced damage to surrounding healthy tissues (sparing effect) not compromising the integral antitumor effectiveness. As evidenced by many reviews, \cite{farr2022,schulte2023,limoli2023,chow2024,vozenin2024,rosini2025,feng2025} the underlying UHDR mechanisms remain elusive. Here we develop a quantitative physical model of UHDR effects. This work is limited to basic physics ignoring biological processes which may be oversimplification. 

Our motivation here is driven by the observations that healthy tissues remain perceptive to radiation damages for dose rates below a certain threshold $\dot{D}_c$. However, they become much less responsive when dose rate $\dot{D}>\dot{D}_c$. That observation can be phenomenologically interpreted as originating from some collective behavior characteristic of materials with high concentration of plasma (positive and negative charges). The only such phenomenon known to us is that of the electron-hole liquid (EHL), previously known with perfect semiconductor crystals at low temperatures \cite{keldysh1970,pokrovskii1972,tikhodeev1985,keldysh1986,ogawa1986,shimano2002,nagai2003,sauer2004,almand2014,poonia2023} and never discussed for water at room temperature or likewise objects. 

The nontrivial features of EHL in the preceding research are essentially quantum implying the non-localized electrons and holes in their corresponding bands of ideal crystalline structures. For such a `quantum EHL', there exists a critical temperature $T_c$, below which EHL remains stable. With a few exceptions, $T_c\sim 100$ K. The `quantum EHL' cases reveal themselves in the observed optical spectra, magnetism, and electrical conduction. For example, \cite{nagai2003} EHL photoluminescence emerges much later than that of stand-alone charge carrier excitations, which reflects the kinetics of EHL condensation. 

\begin{figure}[b!]
\includegraphics[width=0.47\textwidth]{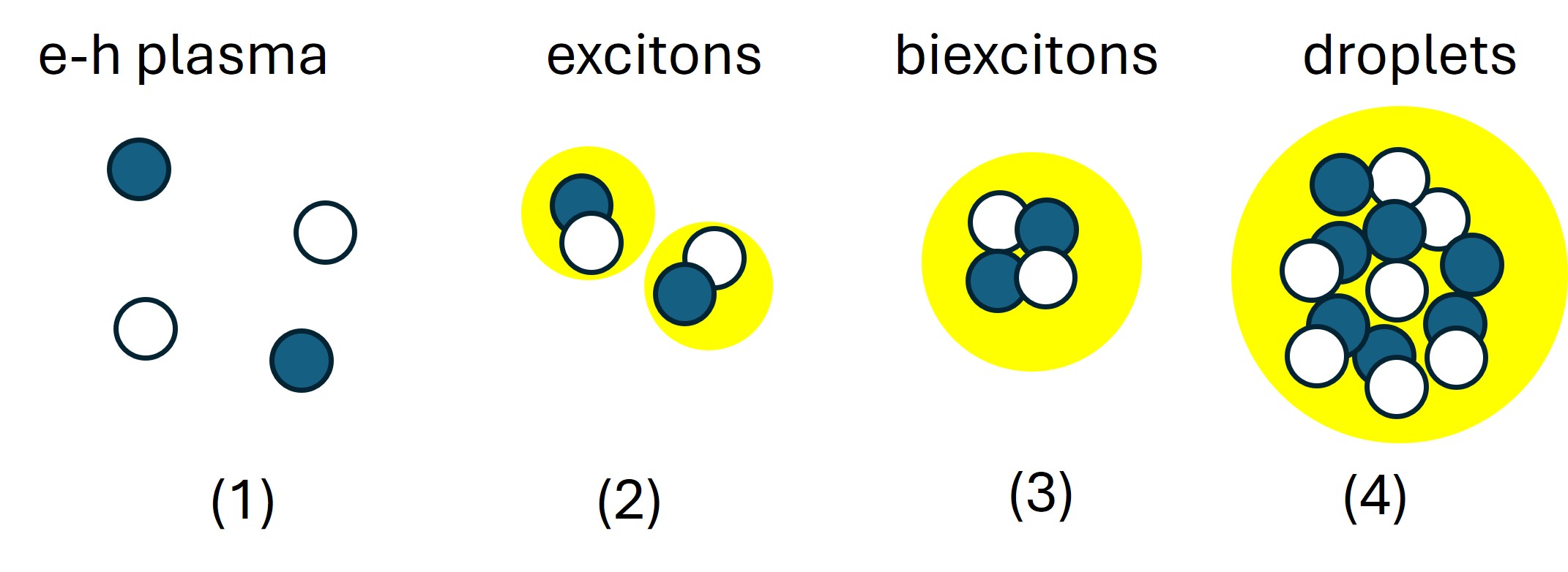}           
\caption{Radiation induced ionization depending on dose rate and energy. Dark (white) circles represent electrons (holes). From left to right: (1) plasma of weakly interacting electrons and holes generated with low dose rates high energy radiation; (b) excitons, i. e. coupled electron -
hole pairs generated with higher dose rates and close to the absorption edge radiation; (3) biexcitons which are coupled pairs of excitons forming with increase of exciton concentration; (4) a droplet of electron hole liquid under high dose rate of well absorbed radiation.}\label{Fig:droplet}
\end{figure}

Here, we extend the conjecture of EHL to biological tissues. Our thinking is based on a simple intuitive argument illustrated in Fig. \ref{Fig:droplet} where increase in dose rates leads to qualitative changes in the ionized radiation product towards denser e-h conglomerate, through excitons, then biexcitons, to EHL.

Our paper is organized as follows. In Section \ref{sec:dose} we discuss general property of our EHL and its dose related criterion assuming zero electron-hole recombination. The recombination effects are introduced in Section \ref{sec:rec} along with the corresponding EHL kinetics. Section \ref{sec:disp} discusses EHL effects related to the frequency dispersion of dielectric permittivity that are typical of water based substances and appear important for here introduced EHL. Finally, Section \ref{sec:concl} summarizes our results and their correspondence to the available observations. 

\section{EHL without recombination}\label{sec:dose}

Unlike the `quantum EHL', charge carriers in biological substances 
strongly interact with atomic subsystems creating local atomic rearrangements, forming polarons or solvated electrons in a matter of picoseconds. \cite{emin2013,ghosh2020,bombile2018,conwell2020} Because they carry clouds of atomic deformations, their diffusivities are extremely low, making them qualitatively similar to ionic melts. Following the standard description of ionic systems, the binding energy in such biological EHL is given by,
\begin{equation}\label{eq:madelung} U= \frac{\mathfrak{M}e^2}{\kappa r}.\end{equation}
Here, $\kappa$ is the dielectric permittivity, $r$ is the nearest neighbor distance, and $\mathfrak{M}$ is the Madelung constant \cite{kittel1996,wikiMad,takenaka2024} accounting for interactions with charges beyond the nearest neighbors. Typically, $\mathfrak{M}\sim 2$ varying between different structures. However, it was argued based on both the experimental data and theoretical analysis that much higher values $\mathfrak{M}\gtrsim 10$ are possible in liquid systems. For definiteness, we will assume $\mathfrak{M}=2$ in our numerical estimates below.

To establish the EHL criterion, it is natural to introduce a dimensionless parameter 
\begin{equation}\label{eq:liquid} L=\frac{U}{k_BT}=\frac{\mathfrak{M}e^2}{\kappa rk_BT}\end{equation}
where $k_B$ is the Boltzmann's constant and $T$ is the absolute temperature; $k_BT=0.026$ eV for room temperature $T=300$ K. Depending on its $L$ value, a e-h system can behave as a low concentration plasma ($L\ll 1$), or EHL ($L\gtrsim 1$), or ionic crystal ($L\gg 1$). 

The significance of EHL concept here is that the binding energy barrier $U$ suppresses recombination between the negative and positive components (say, between hydroxide ions and hydronium cations) by making it necessary to first decouple such a pair from the rest of EHL before they can recombine. The probability of such extraction process is estimated in terms of the Boltzmann's exponent, 
\begin{equation}\label{eq:temp}p=\exp(-U/k_BT)=\exp(-L).\end{equation} 
Such a coupling is consistent with the observation of bound pairs of solvated electrons and hydronium cations. \cite{ma2014} 

Another important interpretation is that $U$ presents an increase in diffusion barrier due to Coulomb interaction in EHL. Because of diffusion slowdown, the rate of chemical reactions between the EHL charge carriers will decrease. According to the established understanding, it is due to the latter reactions that a biological system under radiation creates reactive secondary species (free radicals, etc.) capable of killing the living cells \cite{cohen1999} that can be either useful (killing cancer) or detrimental (killing healthy cells). Therefore, emergence of EHL can  
suppress biological activity leading to the sparing effect. Note that suppression of diffusivity in liquids (compared to their corresponding gases) is typical of classical systems; hence, our conclusion about the EHL related sparing is not unexpected.

To estimate $L$ we start with the integral radiation dose $D$. Assuming that ionized charge carriers do not recombine yields their ultimate concentration, 
\begin{equation}\label{eq:conc}n=2D\rho /I,\end{equation} 
in units 1/cm$^3$ where $\rho \approx 1 $ g/cm$^{3}$ is the density of water as the main tissue component, $I$ is the energy of e-h pair creation, and the coefficient 2 accounts for two charge carriers per pair. (Effects of recombination are described next.) The particle concentration and the average distance $r$ between them are related through the standard rigid sphere approximation,
\begin{equation}\label{eq:n-r}n(4\pi r^3/3)=1.\end{equation} Substituting here the above $n$ yields
\begin{equation}\label{eq:distance} r=\left(\frac{3I}{8\pi D\rho}\right)^{1/3}.\end{equation}
As a numerical example, the average nearest neighbor distance in EHL is estimated as 7 nm assuming $D=30$ Gy (escalated over the standard regime towards UHDR values).

Using the latter $r$ in Eq. (\ref{eq:madelung}) yields the value of parameter $L$. Setting then $L>1$ defines a condition on integral dose under which EHL can form,
\begin{equation}\label{eq:flash}D>D_{\rm min}=\left(\frac{\kappa k_BT}{\mathfrak{M}e^2}\right)^3\frac{3I}{8\pi\rho}.\end{equation}
For numerical estimates we set $I=10$ eV, and $\kappa\approx 5$ based on the published data for fat in human tissues \cite{tannino2023} and in bacterial cell. \cite{checa2019} The choice for $\kappa$ will be discussed more in detail in Sec. \ref{sec:disp} below. Substituting these values yields $D_{\rm min}\sim 10$ Gy, in the ballpark of conventional radiation treatment doses.\cite{farr2022,schulte2023,limoli2023,chow2024,vozenin2024,rosini2025}

\section{EHL with recombination}\label{sec:rec}
While the criterion of Eq. (\ref{eq:flash}) is given in terms of dose, it is a general perception that the UHDR effects are defined by the dose rates $\dot{D}$. Our approach here is that the dose of Eq. (\ref{eq:flash}) can be expressed in terms of $\dot{D}$ when recombination in EHL is taken into account.

Tacitly assuming zero recombination in deriving Eq. (\ref{eq:flash}) implies that, in the very beginning of radiation exposure, the recombination time $\tau _0$ is long enough to allow charge accumulation rather than depletion under charge generation rate $g$ related to the dose rate as $g=2\rho\dot{D}/I$. Following the standard kinetic analysis we present the recombination driven decrease in carrier concentration as $n=n_0\exp(-t/\tau _0)$, i. e. recombination rate 
\begin{equation}\label{eq:recomb}(\partial n/\partial t)_{\rm recomb}=-n/\tau _0.\end{equation} 
Including both the generation and recombination processes yields, $dn/dt=g-n/\tau _0$. We thus require $dn/dt>0$, i.e. $g>n/\tau _0$. Using the above expression for $g$ and $n=2\rho D/I$, the latter inequality yields the criterion
\begin{equation}\label{eq:doserate} \frac{\dot{D}}{D}>\frac{1}{\tau _0} ,\end{equation}
which should be applied along with the previously derived criterion $D>D_{\rm min}$ of Eq. (\ref{eq:flash}). Since $\dot{D}/D$ expresses a reciprocal of the dose deposition time, the latter criterion states that the entire dose $D>D_{\rm min}$ must be delivered during a short time not exceeding the initial recombination time in healthy tissues. 

For numerical comparison, we assume water as a major component and, from independent study of electrolysis, take $\tau _0\sim 0.01-1$ s.  \cite{formal2014} Also, as in the above, we set $I=10$ eV, and $\kappa\approx 5$.
Combining the above yields the following criteria for UHDR effects: 
\begin{equation}\label{eq:flash_criteria} D_{\rm min}\sim 10{\rm Gy},\quad \dot{D}_{\rm min}=\frac{D_{\rm min}}{\tau _0}\sim 10-10^3 \frac{{\rm Gy}}{{\rm s}}.\end{equation}

A drawback of Eq. (\ref{eq:recomb}) based approach is that it ignores the charge carrier binding with EHL. Taking the latter into account dictates renormalization $\tau _0\rightarrow p\tau _0$ with $p$ from Eq. (\ref{eq:temp}) and $r$ expressed through $n$ according to Eq. (\ref{eq:n-r}). As a result, Eq. (\ref{eq:recomb}) takes the form of a kinetic equation,
\begin{equation}\label{eq:kinetics}\frac{dn}{dt}=-\frac{n}{\tau _0}\exp\left[-\frac{\mathfrak{M}e^2}{\kappa k_BT}\left(\frac{4\pi n}{3}\right)^{1/3}\right]+g\end{equation}
where we have added the charge generation rate $g$.

\begin{figure}[bht] 
\includegraphics[width=0.4\textwidth]{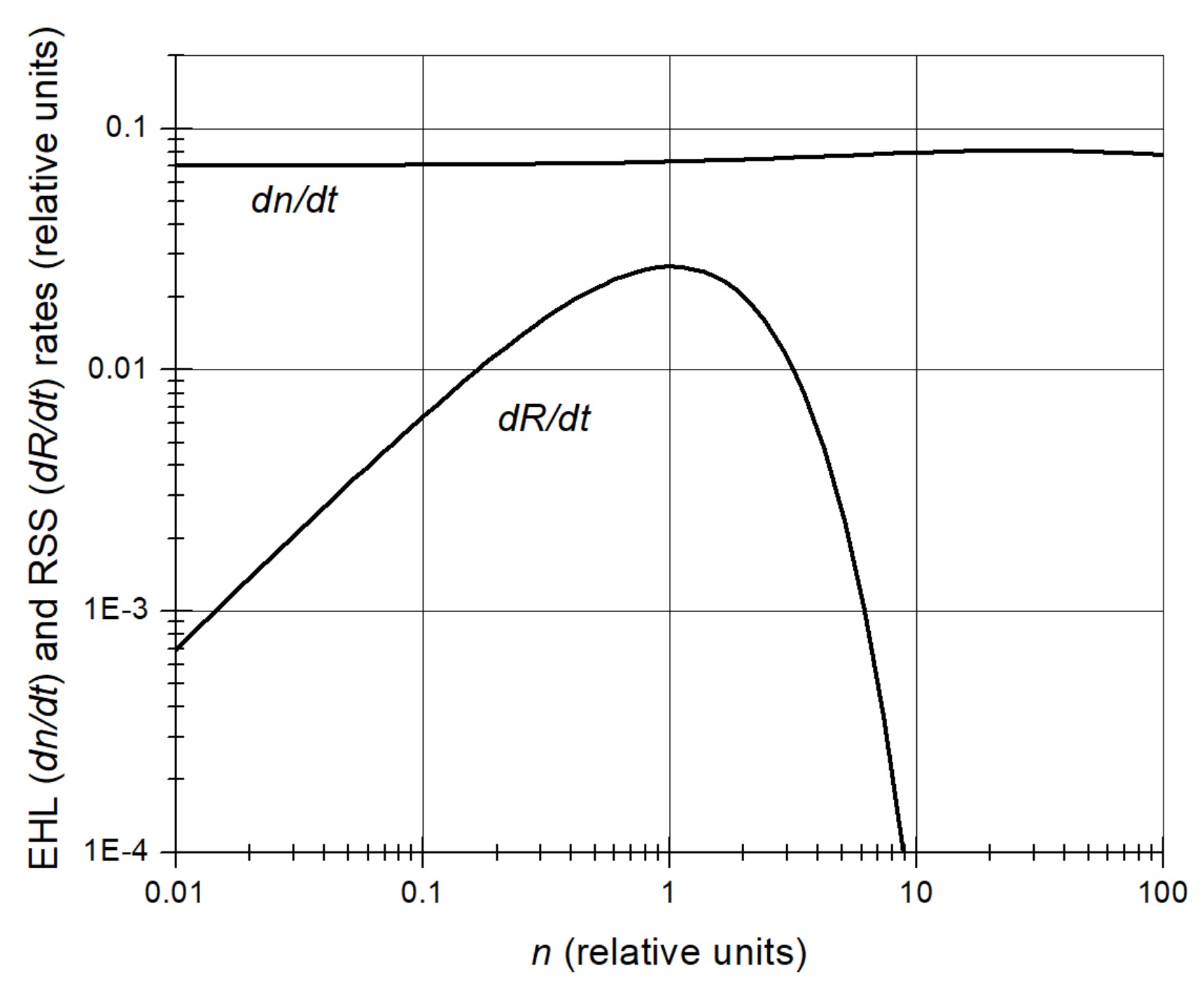} 
\caption{Concentration dependencies of EHL ($dn/dt$) and reactive secondary species ($dR/dt$) rates assuming strong enough binding ($L>2$). }\label{Fig:FLASH_rates}
\end{figure}

According to our theory, the reactive species are generated by charge carriers that overcome EHL binding. Their concentration ($R$) is given by  
\begin{equation}\label{eq:RSS}R(t)=\int _0^t\frac{n}{\tau _0}\exp\left[-\frac{\mathfrak{M}e^2}{\kappa k_BT}\left(\frac{4\pi n}{3}\right)^{1/3}\right]dt.\end{equation}
Depicted in Fig. \ref{Fig:FLASH_rates} are the rates of EHL concentration $dn/dt$ and the concentration $dR/dt$ of charges departed from EHL that are capable of forming reactive secondary species (radicals, etc.).
We observe that $dR(t)/dt$ first increases as EHL accumulates, and then decreases when EHL binding becomes strong enough to suppress the charge carrier departure. 

\begin{figure}[h!]
\includegraphics[width=0.42\textwidth]{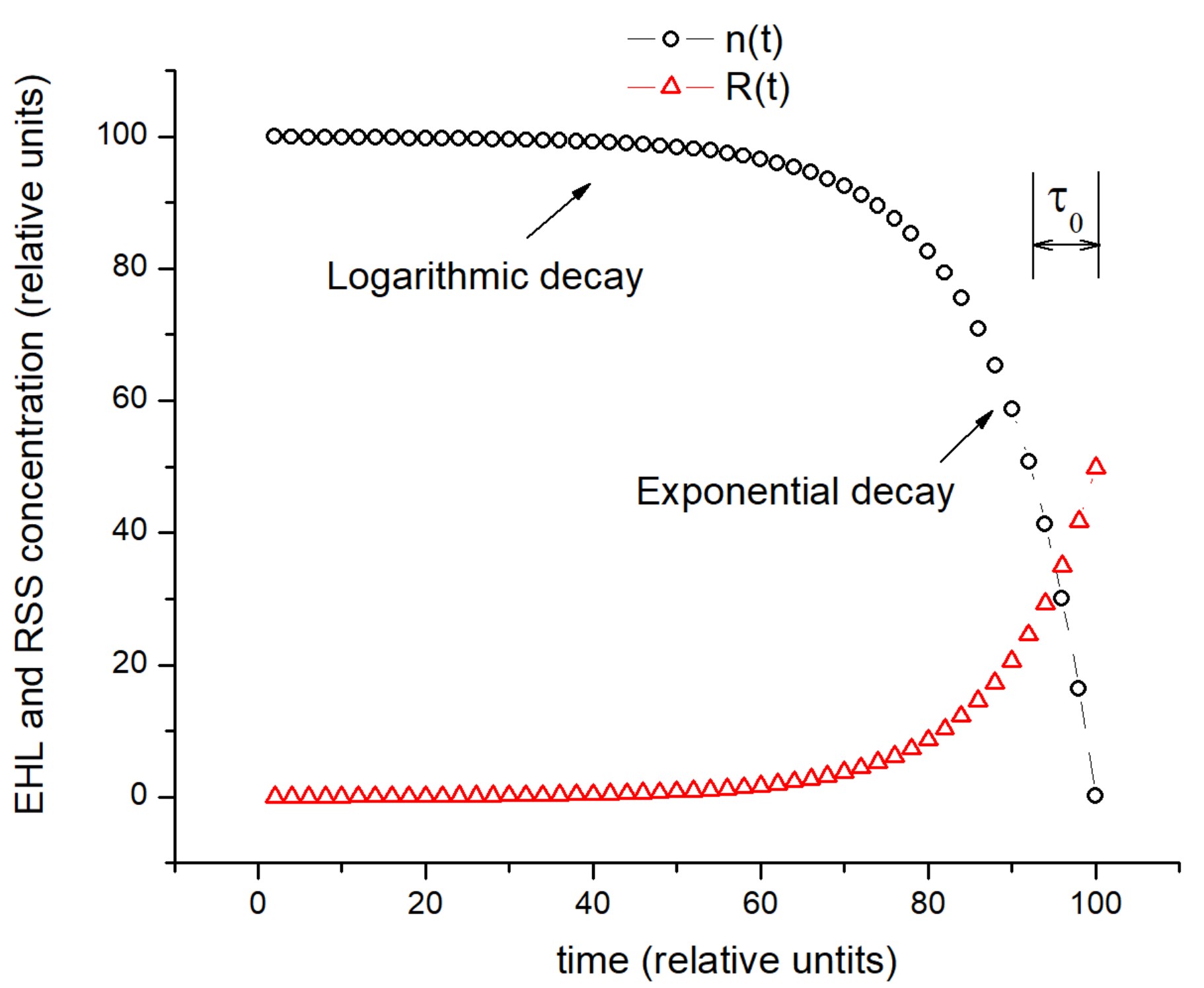}           
\caption{A sketch of temporal dependencies for EHL and RSS concentrations with initial strong binding, $L>2$ upon switching off the radiation source. It is assumed that creation of each RSS particle requires two charge carriers from EHL.  }\label{Fig:nt}
\end{figure}

Once EHL is formed, it persists being almost intact for a certain time interval after the radiation is turned off (making $g=0$ in Eq. (\ref{eq:kinetics}))because the EHL binding remains about the same during several $\tau _0$. Such a process is presented in Fig. \ref{Fig:nt} as the integral of Eq. (\ref{eq:kinetics}) with $g=0$.

The decay of Fig. \ref{Fig:nt} reveals first a degree of EHL stability due to the mutual binding of charges in EHL resulting in a relatively weak temporal dependence close to logarithmic. However that decay accelerates becoming exponential after the EHL concentration decreases enough to considerably weaken the binding.  The latter prediction can be useful for identifying EHL effects experimentally, through optical spectra triggered by UHDR exposure, similar to the experiments identifying solvated electrons. \cite{kenee1960,hart1963}  

\begin{figure}[bht] 
\includegraphics[width=0.35\textwidth]{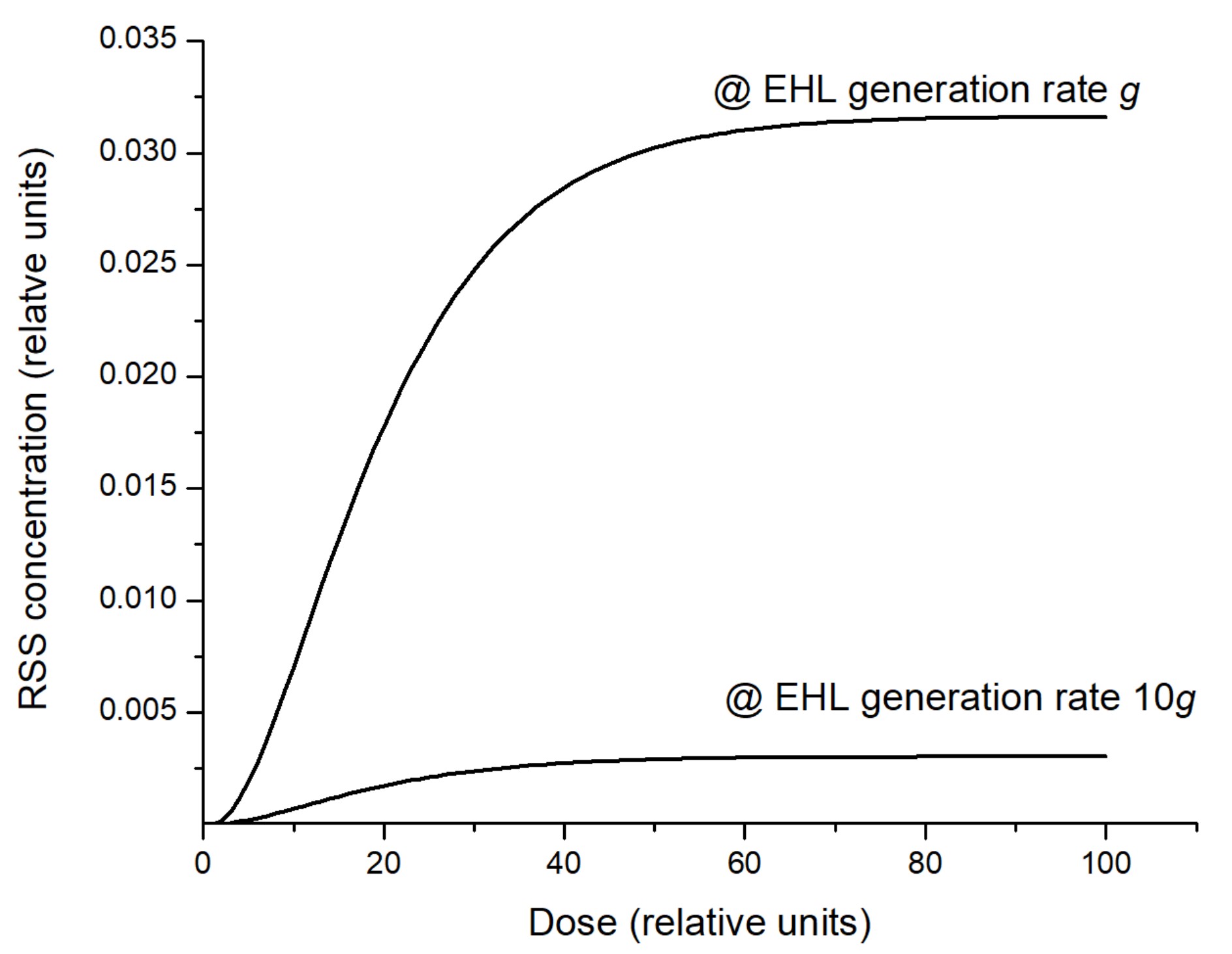} 
\caption{A calculated sketch of sparing effect showing how irradiation created modifications {\it decrease} with the dose rate. }\label{Fig:SPLASH_spare}
\end{figure}

Also, following Eq. (\ref{eq:kinetics}) we were able to verify the biological sparing effect, according  to which radiation inflicted damages decrease with the dose rate. Such evidence of the sparing effect is shown in Fig. \ref{Fig:SPLASH_spare} presenting $R(t)$ from Eq. (\ref{eq:RSS}) corresponding to an order of magnitude different generation rates, $g$ and $10g$. These curves are qualitatively similar to recently published experimental data \cite{gupta2023,debbio2025}. In light of our theory, that sparing effect is due to stronger EHL and diffusion arrest under more intense radiation.


\section{Frequency dispersion effects}\label{sec:disp}

There is yet another constraint defining the minimum dose rate: the time (frequency) dispersion of dielectric permittivity, known as $\alpha$ - dispersion in the low frequency region. \cite{zimmermann2021,matzler1987,elton2016} It originates slow processes of molecular movements, protein reorientations, etc. determining the low frequency dielectric permittivity of water and related substances. These slow relaxation processes make low frequency dielectric permittivity very high, $\kappa\approx 80$. 

With the dielectric constants $\kappa\approx 80$, our parameter $L$ in Eq. (\ref{eq:liquid}) decreases by more than order of magnitude effectively eliminating the EHL coupling. To retain the latter,  $D_{\rm min}$ of Eq. (\ref{eq:flash}) must increase by almost three orders of magnitude vs estimated in the above. However, as pointed out in many cited papers, \cite{zimmermann2021} the slow atomic processes start showing up only at times exceeding 1-10 ms. At shorter time intervals, they do not contribute much, and dielectric permittivity is mostly of electronic nature, about the same as in the above mentioned fat tissues (we note that the dielectric permittivity of ice is close to 3.17. \cite{matzler1987}). Coincidentally, the molecular movement time constraint is of almost the same order of magnitude as our first culprit of recombination, i.e. $\sim 1-10$ ms. Therefore, our estimate in Eq. (\ref{eq:flash_criteria}) remains qualitatively the same. 

Because the `frozen' low frequency molecular dynamic can be an important part of UHDR effects, we provide in this section its approximate analysis based on Eq. (\ref{eq:kinetics}) with $\kappa$ being a time dependent variable. It is convenient to use its simple model in the form, 
\begin{equation}\label{eq:z}z(t)\equiv \kappa ^{-1}=\kappa _0^{-1}\exp(t/\tau _{\kappa})\end{equation}
where $\tau _\kappa \sim 1-10$ ms plays the role of a material parameter and $\kappa _0$ will not show up in our calculations.

We then describe the kinetics in terms of independent variable $z$ noting that 
\begin{equation}\label{eq:dndt}\frac{dn}{dt}=\left(\frac{z}{\tau _\kappa}\right)\frac{dn}{dz}.\end{equation}
Introducing a new function
\begin{equation}\label{eq:u}u= z\left[\frac{\mathfrak{M}e^2}{\kappa k_BT}\left(\frac{4\pi n}{3}\right)^{1/3}\right], \end{equation}
Eq. (\ref{eq:kinetics}) takes the form
\begin{equation}\label{eq:kinetics1}\frac{3}{\tau _\kappa}\left(\frac{du}{dz}-\frac{u}{z}\right)=\frac{u}{z\tau _0}\exp(-u).
\end{equation}
\begin{figure}[b!]
\includegraphics[width=0.45\textwidth]{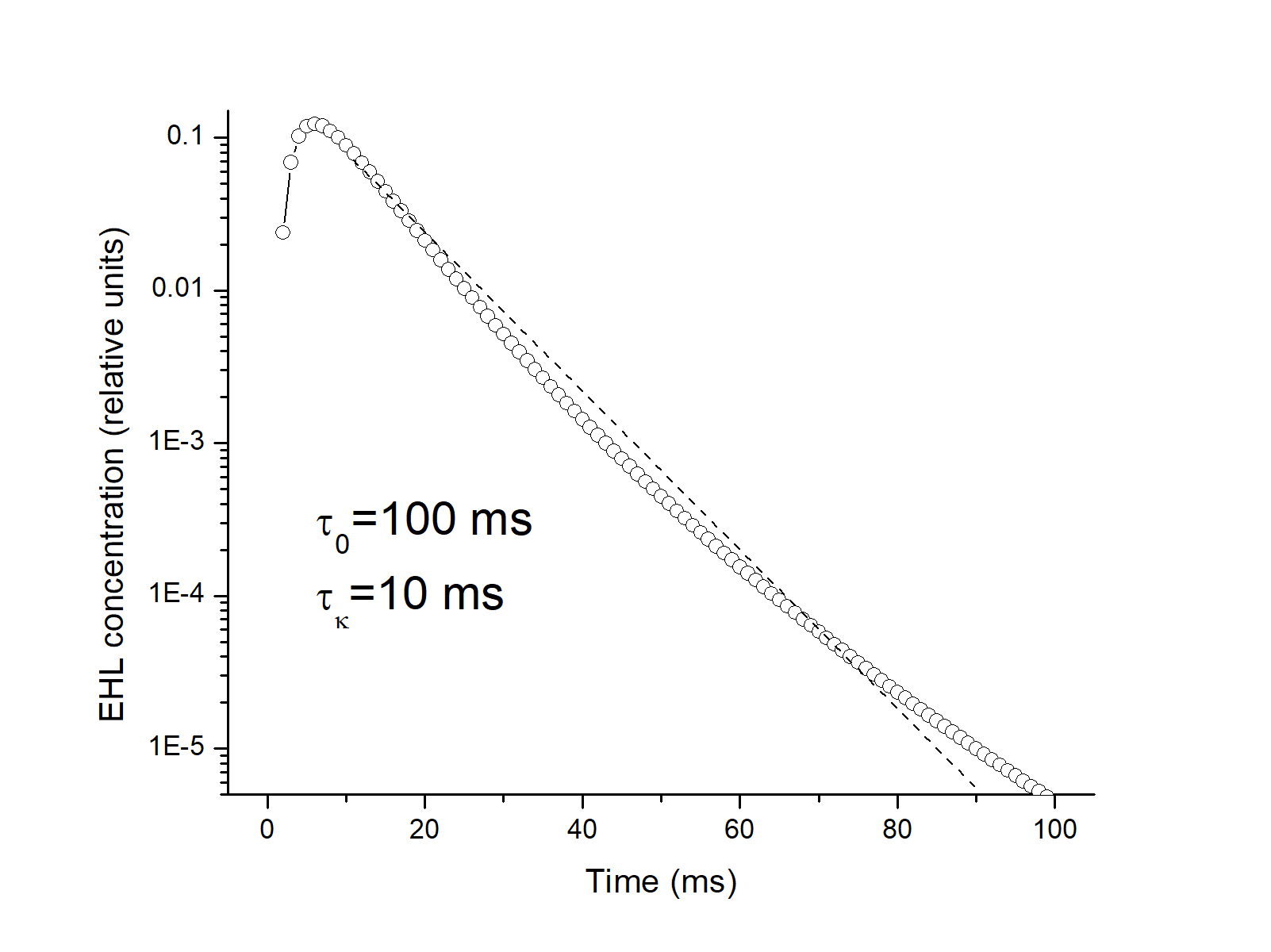}           
\caption{A temporal dependence of EHL concentration $n$ for  recombination and dielectric permittivity relaxation processes acting simultaneously. The corresponding times in the figure will serve as our example. The maximum point in the curve reflects increase in EHL concentration before the dielectric relaxation starts. Dashed line represents an approximation of the purely exponential decay $\exp(-t/\tau _{\kappa})$. }\label{Fig:EHL_disp}
\end{figure}

Its solution can be represented as,
\begin{equation}\label{eq:kinsol} \int _{u_0}^u\frac{\tau _0\exp(u)du}{u[3\tau _0\exp(u)+\tau _\kappa ]} =\ln \left[\frac{z(t)}{z(0)}\right] +C\end{equation}
with integration constant $C$ determined by the initial condition $u(t=0)=u_0$.

Assuming the temporal model of Eq. (\ref{eq:z}), the right-hand-side in Eq. (\ref{eq:kinsol}) reduces to simply $t/\tau _\kappa$. Its left-hand-side, while cannot be expressed in elementary functions, allows a quick numerical solution illustrated in Fig. \ref{Fig:EHL_disp}. Comparing Figs. \ref{Fig:nt} and \ref{Fig:EHL_disp} it is obvious that the frequency dispersion of dielectric permittivity can significantly affect the temporal dependence of EHL decay. 

\section{Discussion and conclusions}\label{sec:concl}
Our theory is limited to physical processes thus predicting UHDR effects at the cellular and subcellular levels of hierarchy, much below that of organisms. In that connection, we would like to mention a recent publication \cite{gupta2023} experimentally evaluating dose rate effects on oxidative damage to cellular culture. It was found that the higher dose rates result in less damage, although no underlying physical mechanism was determined. Similarly, a recent publication \cite{debbio2025} explored the molecular basis of electron UHDR effect in healthy human bronchial epithelial cells with the conclusion that it resulted in lower RSS production.  Phenomenologically the EHL based mechanism of sparing effect correlates with the "free radical hypothesis" \cite{feng2025} and hypothetical scavenger effects, \cite{alhaddad2024} according to which some undefined processes diminish the concentration of free radicals. Our paper attributes such processes to EHL binding.

We note that our `kinetic' results presented in figures \ref{Fig:FLASH_rates}, \ref{Fig:nt}, \ref{Fig:SPLASH_spare}, and  \ref{Fig:EHL_disp} can be rather sensitive to parameter  choice. The depicted trends are aimed at illustrating the major trends only; more accurate tuning  (possibly taking into account some other kinetic processes) will be needed to compare our predictions with actual data. 

Our theory predicts that EHL effects are not expected within cancer morphologies where their inherent  disorder shortens recombination times by many orders of magnitude \cite{shen1994,baranovskii1988} killing EHL.  Lacking EHL, free radical generation and cancer killing  effects are expected to prevail. 


%
\section*{acknowledgements} D. S. would like to thank Drs. Eunsin Lee, Ashley Cetnar, Ahmet Ayan, Wei Meng, Nilendu Gupta, and Arnab Chakravarti for many enthralling discussions.

\end{document}